\documentclass[twocolumn,showpacs,preprintnumbers,amsmath,amssymb,superscriptaddress]{revtex4}

\usepackage{amsmath,amsfonts,amssymb}
\usepackage[english]{babel} 
\usepackage[latin1]{inputenc} 
\usepackage[T1]{fontenc}
\usepackage{color}
\usepackage{float}
\usepackage{verbatim}
\usepackage{graphicx}
\usepackage{bm}

\def \equi#1{\mathrel{\mathop{\kern 0pt\sim}\limits_{#1}}} 
\newcommand{\deriv}[2]{\frac{\mathrm{d}#1}{\mathrm{d}#2}}

\newcommand{\diff}[1]{\mathrm{d}#1 \;}
\newcommand{\e}[1]{\; \mathrm{e}^{#1} \;}
\newcommand{\erfc}[1]{\; \mathrm{erfc}\left(#1\right) \;}
\newcommand{\erf}[1]{\; \mathrm{erf}\left(#1\right) \;}
\newcommand{\interval}[1]{\ensuremath{\left[#1\right]}}

\begin{document}

\title{Survival probability of a Brownian motion in a planar wedge of arbitrary angle}

\author{Marie Chupeau}
\affiliation{Laboratoire de Physique Th\'eorique de la Mati\`ere Condens\'ee (UMR CNRS 7600), Universit\'e Pierre et Marie Curie, 4 Place Jussieu, 75255
Paris Cedex France}

\date{\today}
\author{Olivier B\'enichou}
\affiliation{Laboratoire de Physique Th\'eorique de la Mati\`ere Condens\'ee (UMR CNRS 7600), Universit\'e Pierre et Marie Curie, 4 Place Jussieu, 75255
Paris Cedex France}

\author{Satya N. Majumdar}
\affiliation{CNRS, Laboratoire de Physique Th\'eorique et Mod\`eles Statistiques, Univ. Paris-Sud, 91405 Orsay Cedex, France}

\begin{abstract}
We study the survival probability and the first-passage time distribution for a Brownian motion in a planar wedge with infinite absorbing edges. We generalize existing results obtained for wedge angles of the form $\pi/n$ with $n$ a positive integer to arbitrary angles, which in particular cover the case of obtuse angles. We give explicit and simple expressions of the survival probability and the first-passage time distribution in which the difference between an arbitrary angle and a submultiple of $\pi$ is contained in three additional terms. As an application, we obtain the short time development of the survival probability in a wedge of arbitrary angle.
\end{abstract}

\maketitle

\section{Introduction}
 
 The survival probability, which is the probability not to have reached a target up to a given time, is a key observable in the study of Brownian motion. It is the cumulative of the first-passage time distribution, \textit{i.e.} the distribution of the time to reach a target. This quantity \cite{Redner,Majumdar:1999a,Bray:2013,BenichouNatChem10,Meyer:2011}, and as a first step the mean first-passage time \cite{CondaminNature07}, is a standard way to quantify the efficiency of a search process. It is for example also involved in the calculation of the covered territory \cite{Hughes,Hilhorst91} and the mean perimeter of the convex hull of a Brownian motion \cite{Majumdar:2010}. 
 
 Determining the survival probability and the first-passage time distribution is of particular importance in the wedge geometry. Recent related studies include last-passage times determination \cite{Desbois:2003}, extensions to anomalous diffusion \cite{Lenzi:2009} (and in particular to fractional Brownian motion \cite{Metzler:2011}) and applications to virus trafficking in cells \cite{Lagache:2008}. 
 Interest in this geometry resides in part in the possibility to map one-dimensional diffusion-controlled reaction processes on a wedge domain \cite{Redner,Redner:1999,Fisher:1988}. One important example of this mapping is the Fisher-Gelfand three-particle problem \cite{Fisher:1988,Bray:2013,Redner}. Consider three diffusing particles on a line, with diffusion constants $D_1$, $D_2$ and $D_3$. Given the starting positions $x_1$, $x_2$ and $x_3$, what is the survival probability of the middle particle up to time $t$, that is to say the probability that it has not met the two other particles up to time $t$? By writing the Fokker-Planck equation in coordinates $y_1=x_1-x_2$ and $y_2=x_2-x_3$, and after some transformations, this problem reduces precisely to the problem of the survival probability of a single Brownian motion in a 2D wedge with top angle $\alpha= 2 \arctan[\sqrt{(1-\gamma)/(1+\gamma)}]$ where $\gamma=D_2/\sqrt{(D_1+D_2)(D_2+D_3)}$. 
 
 In the general case, the survival probability for regular diffusion in a wedge domain is written as an infinite sum of special functions. In the past, special attention has been devoted to the analysis of the large time behavior of the survival probability, which displays a power law decay with an exponent continuously depending on the wedge angle \cite{Redner}. On the other hand, the analysis of the short time behavior seems to have been left aside.
 
 Recently, and in contrast with the standard expression involving an infinite sum, compact analytical expressions of the survival probability and the first-passage time distribution have been obtained for special values of the wedge angle by using the method of images \cite{Dy:2008}. These expressions have been extended to biased diffusion \cite{Dy:2013}. However, these results have been limited to specific wedge angles of the form $\pi/n$ where $n$ is a positive integer. In particular, they do not apply to obtuse angles.

 Here, focusing on unbiased diffusion, (i) we generalize these results to \textit{arbitrary} wedge angles; Note that this covers in particular the case of obtuse angles, which has proven to be an essential ingredient in the context of diffusion growth processes \cite{Redner,Turkevich:1985}; Beyond this, it is crucial to know the survival probability in a wedge of arbitrary angle to solve the Fisher-Gelfand problem presented above for general diffusion coefficients;
 (ii) The expressions presented here take the form of a \textit{finite} sum over generalized images plus an integral only involving \textit{elementary} functions, which vanishes for wedge angles of the form $\pi/(2p+1)$; This structure thus explicitly underlines the difference between a wedge angle $\pi/n$ and an arbitrary one;
 (iii) As an application, we show that the short time behavior of the survival probability is conveniently obtained from this expression.
 
The manuscript is organized as follows. In Sec. II, we remind the standard expression of the survival probability and give the main results of this paper. In Sec. III, we provide the derivation of the alternative expressions of the survival probability and the first-passage time distribution for any wedge angle - the most technical part of the derivation appears in appendix - and compare them with the existing results for special $\pi/n$ wedge angles. In Sec. IV, we give the short time asymptotic development of the survival probability. Finally, in Sec. V, we draw our conclusions.

\section{Basic equations and main results}
\begin{figure}[h]
\centering
\includegraphics[width=100pt]{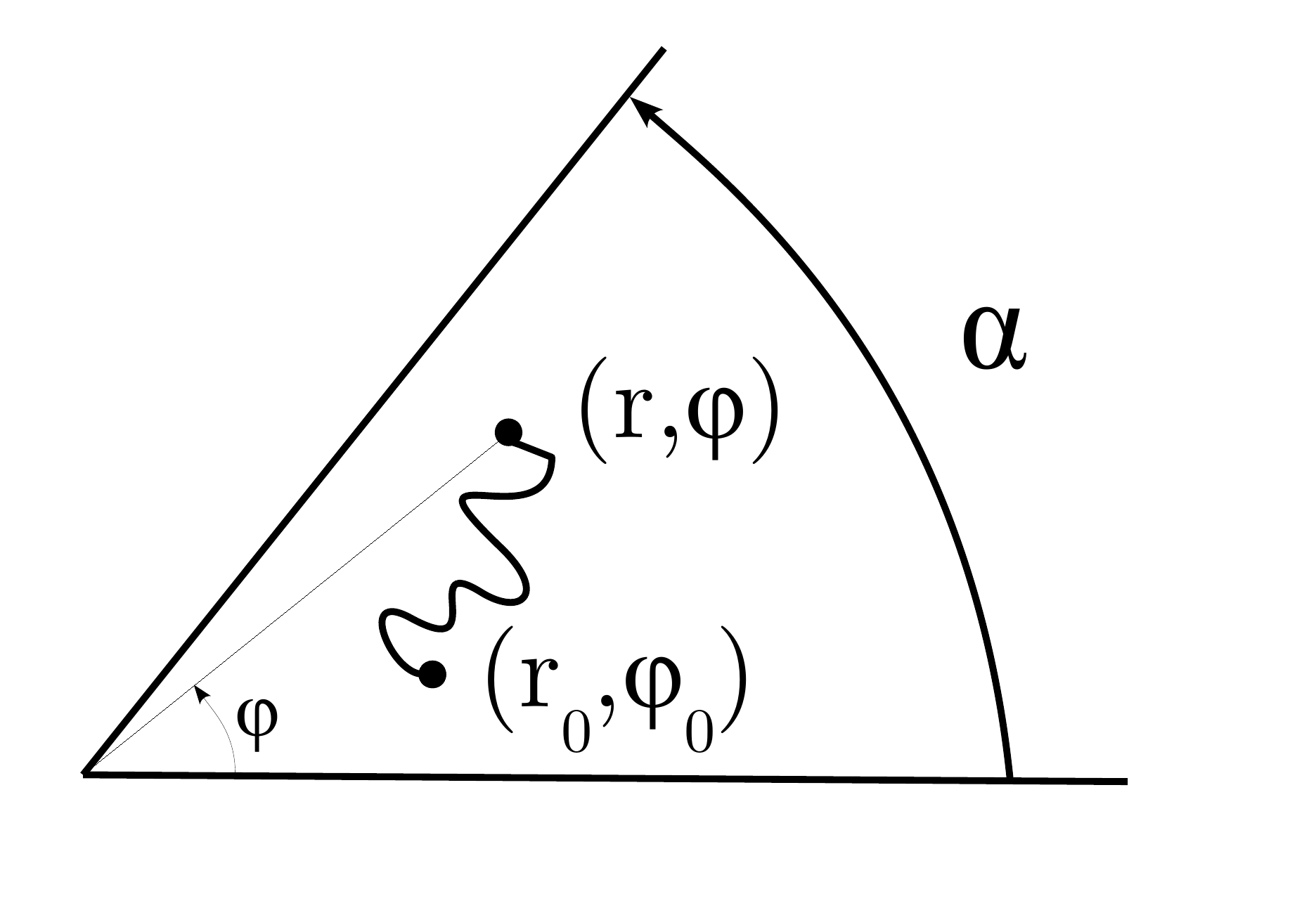}
\caption{Geometry of the wedge. The starting point is $(r_0,\varphi_0)$ at time $0$ and the position of the walker at time $t$ is $(r,\varphi)$.}
\label{geom}
\end{figure}

Let us consider a Brownian motion in a planar wedge of top angle $\alpha$ with two infinite absorbing edges, starting from a point $(r_0,\varphi_0)$ (see Fig.~\ref{geom}).
The propagator $P(r,\varphi,t|r_0,\varphi_0)$ satisfies the diffusion equation
\begin{equation}
\dfrac{\partial P}{\partial t}=D \Delta P= D \left[ \dfrac{\partial^2 P}{\partial r^2} + \dfrac{1}{r}\dfrac{\partial P}{\partial r}+ \dfrac{1}{r^2} \dfrac{\partial^2 P}{\partial \varphi^2} \right]
\end{equation}
with the following initial condition and boundary conditions:
\begin{eqnarray} \label{backward}
&&P(r,\varphi,0|r_0,\varphi_0)=\delta(\bm{r}-\bm{r_0})=\dfrac{1}{r_0} \delta(r-r_0) \delta(\varphi-\varphi_0) \nonumber \\
&&P(r,0,t|r_0,\varphi_0)=P(r,\alpha,t|r_0,\varphi_0)=0
\end{eqnarray}
The exact solution of this problem is known \cite{Desbois:2003,Lagache:2008}
\begin{eqnarray}
&P(r,\varphi,t|r_0,\varphi_0)&=\dfrac{1}{\alpha D t} \sum\limits_{n=1}^{+\infty} \sin\left(\frac{n \pi \varphi}{\alpha} \right) \sin\left(\frac{n \pi \varphi_0}{\alpha} \right)  \nonumber \\ 
&& \times \, \mathrm{I}_{\frac{n\pi}{\alpha}}\left(\frac{r_0 r}{2Dt} \right) \exp \left(- \frac{r^2+r_0^2}{4Dt} \right).
\end{eqnarray}
The survival probability, defined by
\begin{equation}
S(t|r_0,\varphi_0)=\int_0^{+\infty} \diff{r} r   \int_0^{\alpha} \diff{\varphi} P(t,r,\varphi|r_0,\varphi_0), 
\end{equation}
can be computed by integration by parts, as a function of a rescaled variable $y=r_0^2/(8Dt)$ and the initial angle $\varphi_0$
\begin{eqnarray}\label{surviey}
&S(y,\varphi_0) = &2 \; \sqrt{\frac{2y}{\pi}} \; \mathrm{e}^{-y} \sum\limits_{m=0}^{+\infty} \frac{\sin\left((2m+1)\frac{\varphi_0 \pi}{\alpha}\right)}{2m+1} \nonumber \\
&& \quad \times \left[ \mathrm{I}_{\nu}(y)+\mathrm{I}_{\nu+1}(y) \right]
\end{eqnarray}
with \mbox{$\nu=(2m+1)\pi/(2\alpha)-1/2$}.

While this expression is well-suited to analyze the large time (small $y$) behavior of the survival probability \mbox{$S(y,\varphi_0) \underset{y \to 0}{\propto} y^{\pi/(2\alpha)}$}, it is difficult to extract the small time (large $y$) behavior. Indeed, using bluntly the behavior of the modified Bessel function to large argument
\begin{equation}
\mathrm{I}_{\nu}(y) \equi{y \to\infty} \frac{\e{y}}{\sqrt{2\pi y}} \left( 1- \frac{4\nu^2-1}{8y}\right)
\end{equation}
leads to \mbox{$S(y,\varphi_0)\underset{y\to\infty}{\longrightarrow}1$} but does not allow to obtain higher order corrections. The main purpose of this paper is to provide this small time (large $y$) asymptotics. We proceed to provide a compact alternative expression of the survival probability (summarized in Eqs.~\eqref{acute} and \eqref{obtuse}) that is more suitable for the large $y$ analysis. The result for the large $y$ asymptotics are summarized in Eqs.~\eqref{asymptotics}, \eqref{erfscale} and \eqref{remainder}. In particular, we show that for large $y$,
\begin{equation}
S(y,\varphi_0) \equi{y\to\infty} 1-\erfc{\sqrt{2y} \min \left(\varphi_0,\frac{\pi}{2} \right)}
\end{equation}
and also provide further subleading corrections.

\section{Derivation of the survival probability and the first-passage time distribution}

The starting point to establish this alternative expression of the survival probability is the integral representation of the modified Bessel function I$_{\nu}$
\begin{eqnarray}
&\mathrm{I}_{\nu}(y)=&\frac{1}{\pi} \int_0^{\pi} \mathrm{d} \varphi \; \mathrm{e}^{y\cos\varphi} \cos(\nu\varphi) \nonumber \\
&&-\frac{\sin(\nu \pi)}{\pi} \int_0^{+\infty} \mathrm{d}u \; \mathrm{e}^{-y \cosh u -\nu u}.
\end{eqnarray}
Plugging this form into Eq.~\eqref{surviey}, the survival probability becomes
\begin{eqnarray} \label{survieydev}
&&S(y,\varphi_0) = \left(\frac{2}{\pi}\right)^{3/2} \sqrt{y} \; \mathrm{e}^{-y} (A_1+A_2)
\end{eqnarray}
with
\begin{eqnarray}
& A_1&\equiv \int_0^{\pi} \mathrm{d} \varphi \; \mathrm{e}^{y\cos\varphi} \sum\limits_{m=0}^{+\infty} (\cos(\nu\varphi)+\cos((\nu+1)\varphi)) \nonumber \\
&&\quad \times \frac{\sin\left((2m+1)\frac{\varphi_0 \pi}{\alpha}\right)}{2m+1} \nonumber \\
& &= 2 \int_0^{\pi} \mathrm{d} \varphi \; \mathrm{e}^{y\cos\varphi} \cos\frac{\varphi}{2} B_1
\end{eqnarray}
with
\begin{equation}
B_1= \sum\limits_{m=0}^{+\infty} \cos\left((2m+1)\frac{\pi\varphi}{2\alpha}\right) \frac{\sin\left((2m+1)\frac{\varphi_0 \pi}{\alpha}\right)}{2m+1}
\end{equation}
and, replacing $\nu$ with its expression,
\begin{eqnarray}
&& A_2\equiv \int_0^{+\infty} \mathrm{d}u \; \mathrm{e}^{-y\cosh u} \left(\mathrm{e}^{-u}-1\right) \nonumber \\
&&\qquad  \times\sum\limits_{m=0}^{+\infty} \sin(\nu\pi) \frac{\sin\left((2m+1)\frac{\varphi_0 \pi}{\alpha}\right)}{2m+1}  \mathrm{e}^{-\nu u}\nonumber \\
&&\qquad = 2 \int_0^{+\infty} \mathrm{d}u \; \mathrm{e}^{-y\cosh u} \sinh\frac{u}{2} B_2
\end{eqnarray}
with
\begin{equation}
B_2= \sum\limits_{m=0}^{+\infty} \cos\left((2m+1) \frac{\pi^2}{2\alpha}\right) \frac{\sin\left((2m+1)\frac{\varphi_0 \pi}{\alpha}\right)}{2m+1}  \mathrm{e}^{-(2m+1)\frac{\pi u}{2\alpha}}.
\end{equation}

\subsection{Calculation of the term $A_1$}

We show in this section that the term $A_1$ can be explicitly calculated. We first note that the sum $B_1$ can be rewritten as
\begin{equation}\label{B1}
B_1=\int_0^{\frac{\pi \varphi_0}{\alpha}} \mathrm{d}x' \; \sum\limits_{m=0}^{+\infty} \cos((2m+1)x') \cos\left((2m+1)\frac{\pi\varphi}{2\alpha}\right)
\end{equation}
with $\varphi$ varying from $0$ to $\pi$. Then, we make use of the following formula
\begin{equation}\label{dirac}
\sum\limits_{m=0}^{+\infty} \cos\left((2m+1)\frac{\pi y}{L}\right) \cos\left((2m+1)\frac{\pi z}{L}\right) =\frac{L}{4} \delta(y-z),
\end{equation}
 valid for \mbox{$0 \leqslant y\leqslant L/2$} and \mbox{$0 \leqslant z\leqslant L/2$}, for $L=\pi$. Note that special attention must be paid to the ranges of $x'$ and $\pi\varphi/(2\alpha)$ in Eq.~\eqref{B1} in order to use Eq.~\eqref{dirac}. The condition \mbox{$0 \leqslant x' \leqslant \pi/2$} is always respected as \mbox{$\varphi_0 \leqslant \alpha/2$}. In turn, the respect of the condition \mbox{$0 \leqslant \pi\varphi/(2\alpha) \leqslant \pi/2$} for $\varphi$ in $[0,\pi]$ depends on the value of $\alpha$. 
 
 We first address in detail the case \mbox{$\alpha \leqslant \pi$}. The key point is to cut the interval of variation of $\varphi$ into well-chosen intervals on which, after periodicity and parity manipulations on the cosine, the formula (\ref{dirac}) applies. We define the integer $k$ such that 
\begin{equation}\label{defk}
(2k+1)\alpha \leqslant \pi <(2k+3)\alpha,
\end{equation}
that is to say $k=\lfloor\pi/(2\alpha) -1/2\rfloor$. We cut the interval $[0,\pi]$ into three: 
\begin{itemize}
\item the part $[0,\alpha]$,
\item the part $[\alpha,(2k+1)\alpha]$,
\item the part $[(2k+1)\alpha,\pi]$.
\end{itemize}
\begin{figure}[h]
\centering
\includegraphics[width=220pt]{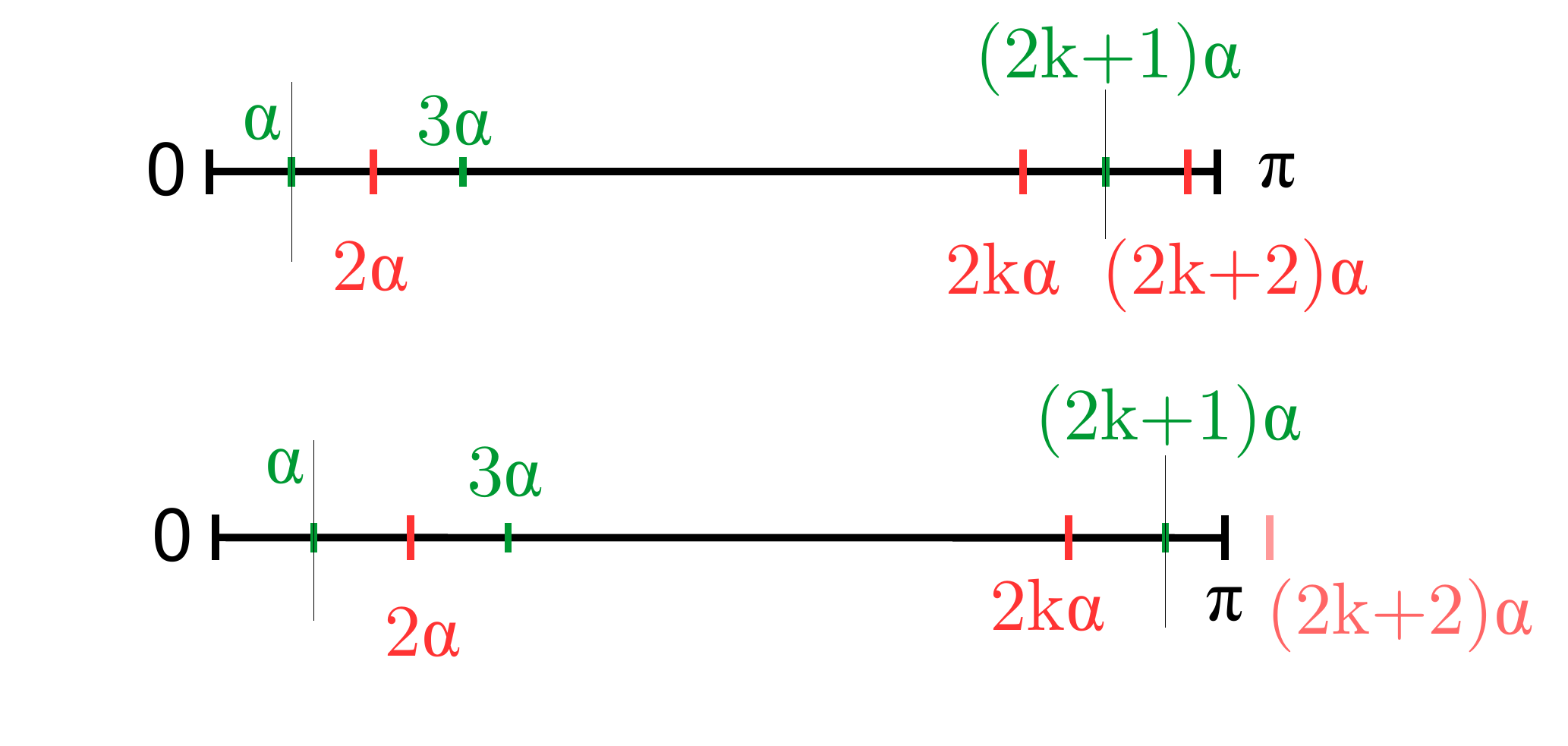}
\caption{The integer $k$ is defined in Eq. (\ref{defk}). The number $(2k+2) \alpha$ can be either smaller than $\pi$ (first case), or greater (second case). In the first case, the interval $[(2k+1)\alpha,\pi]$ has to be cut again into the two intervals $[(2k+1)\alpha,(2k+2)\alpha]$ and $[(2k+2)\alpha,\pi]$.}
\label{decoupage}
\end{figure}

Referring to Fig.~\ref{decoupage}, we can see that two cases arise, depending on whether \mbox{$(2k+2)\alpha$} is smaller or greater than $\pi$. If \mbox{$(2k+2)\alpha<\pi$}, we cut again the last interval into two parts $[(2k+1)\alpha,(2k+2)\alpha]$ and $[(2k+2)\alpha,\pi]$. We can then write the general relation
\begin{eqnarray}
&\displaystyle\int_0^{\pi} \mathrm{d}\varphi \; f(\varphi) =& \underbrace{\int_0^{\alpha} \mathrm{d}\psi \; f(\psi)}_{C_1}  \nonumber \\
&& +\underbrace{\sum\limits_{j=1}^{k} \int_0^{\alpha} \mathrm{d}\psi \; \left(f(2j\alpha-\psi) + f(2j\alpha+\psi)\right)}_{C_2\equiv C_2^-+C_2^+} \nonumber \\
&& +\underbrace{\int_{\max(0,(2k+2)\alpha-\pi)}^{\alpha} \mathrm{d}\psi \; f((2k+2)\alpha-\psi)}_{C_3^-} \nonumber \\
& & +\underbrace{\int_0^{\max(0,\pi-(2k+2)\alpha)} \mathrm{d}\psi \; f((2k+2)\alpha+\psi)}_{C_3^+}. \nonumber \\
\end{eqnarray}
with $C_3^+$ that equals zero in the case where \mbox{$(2k+2)\alpha \geqslant \pi$}. \\

The calculation of the integrals $C_1$, $C_2$, $C_3^-$ and $C_3^+$, carried out in appendix, leads to the final expression for $A_1$
\begin{eqnarray}\label{A1}
&&\kern-1em A_1=\left(\frac{\pi}{2}\right)^{3/2} \frac{\mathrm{e}^y}{\sqrt{y}} \times  \nonumber \\
&&\kern-0.5em \left[ \mathrm{erf}\left(\sqrt{2y}\sin\varphi_0\right)+\sum\limits_{j=1}^{k} (-1)^j   \left[\mathrm{erf}\left(\sqrt{2y}\sin(j\alpha+\varphi_0)\right) \right. \right. \nonumber \\
&& \quad \left. -\mathrm{erf}\left(\sqrt{2y}\sin(j\alpha-\varphi_0)\right)\right] \nonumber \\
&&\quad +(-1)^{k+1}  \left[ \mathrm{erf}\left(\sqrt{2y}\sin\left(\min\left((k+1)\alpha+\varphi_0,\frac{\pi}{2}\right)\right)\right) \right.\nonumber \\
&& \quad  \left. -\, \mathrm{erf}\left(\sqrt{2y}\sin\left(\min\left((k+1)\alpha-\varphi_0,\frac{\pi}{2}\right)\right)\right)\right] \Bigg]. \nonumber \\
\end{eqnarray}
In the case $\alpha \geqslant \pi$, the term $A_1$ is obtained by following the same lines
\begin{equation}
A_1=\left(\frac{\pi}{2}\right)^{3/2} \frac{\mathrm{e}^y}{\sqrt{y}} \erf{\sqrt{2y}\sin\left(\min\left(\varphi_0,\frac{\pi}{2}\right)\right)}.
\end{equation}

\subsection{Calculation of the term $A_2$}

The sum $B_2$ can be rewritten as
\begin{equation}
B_2=\frac{1}{2} \left[ \int_0^{\frac{\varphi_0\pi}{\alpha}} \mathrm{d}x' \; C_2^+(x')+\int_0^{\frac{\varphi_0\pi}{\alpha}} \mathrm{d}x' \; C_2^-(x') \right]
\end{equation}
where
\begin{eqnarray}
&C_2^{\pm}(x')&=\sum\limits_{m=0}^{+\infty}\cos\left((2m+1) \left(x'\pm\frac{\pi^2}{2\alpha} \right) \right) \mathrm{e}^{-(2m+1)\frac{\pi u}{2\alpha}} \nonumber \\
\end{eqnarray}

After simple algebra, we get
\begin{eqnarray}
&C_2^{\pm}(x')&=\mathrm{Re}\left[ \sum\limits_{m=0}^{+\infty} \mathrm{e}^{(2m+1)\left(i\left(x' \pm \frac{\pi^2}{2\alpha}\right)-\frac{\pi u}{2\alpha} \right)} \right] \nonumber \\
&&= \frac{1}{2 \sinh\left(\frac{\pi u}{2\alpha}\right)} \frac{\cos\left(x' \pm \frac{\pi^2}{2\alpha}\right)}{1+\frac{\sin^2\left(x' \pm \frac{\pi^2}{2\alpha}\right)}{\sinh^2\left(\frac{\pi u}{2\alpha}\right)}} .
\end{eqnarray}
We integrate over $x'$
\begin{eqnarray}
&B_2 &=\frac{1}{4} \left[ \arctan\left( \frac{\sin\left(\frac{\pi}{\alpha}\left(\varphi_0+\frac{\pi}{2}\right)\right)}{\sinh\left(\frac{\pi u}{2\alpha}\right)}\right) \right. \nonumber \\
&& \quad \qquad \left.+\arctan\left( \frac{\sin\left(\frac{\pi}{\alpha}\left(\varphi_0-\frac{\pi}{2}\right)\right)}{\sinh\left(\frac{\pi u}{2\alpha}\right)}\right)  \right] 
\end{eqnarray}
and finally,
\begin{eqnarray}
&&\kern-1em A_2 =\frac{1}{2} \int_0^{+\infty} \mathrm{d}u \; \mathrm{e}^{-y\cosh u} \sinh\frac{u}{2}  \times \nonumber \\
&& \kern-1em \left[ \arctan \kern-0.2em \left( \frac{\sin\left(\frac{\pi}{\alpha}\left(\varphi_0+\frac{\pi}{2}\right)\right)}{\sinh\left(\frac{\pi u}{2\alpha}\right)}\right)\kern-0.3em +\arctan\kern-0.2em\left( \frac{\sin\left(\frac{\pi}{\alpha}\left(\varphi_0-\frac{\pi}{2}\right)\right)}{\sinh\left(\frac{\pi u}{2\alpha}\right)}\right)  \kern-0.2em\right]\kern-0.2em. \nonumber \\
\end{eqnarray}

\subsection{Survival probability}

\subsubsection{Expression for an arbitrary angle}

We gather the results of the two previous subsections and give the final expression of the survival probability in an acute wedge of top angle \mbox{$\alpha \leqslant \pi$}
\begin{eqnarray}\label{acute}
&&\kern-1em S(y,\varphi_0) = \erf{\sqrt{2y}\sin\left(\varphi_0 \right)}\nonumber \\
&& +\sum\limits_{j=1}^{k} (-1)^j \left[\erf{\sqrt{2y}\sin\left(j\alpha+\varphi_0\right)} \right.\nonumber \\
&&\qquad  \left. -\erf{\sqrt{2y}\sin\left(j\alpha-\varphi_0\right)} \right]  \nonumber \\
&& +(-1)^{k+1} \left[ \erf{\sqrt{2y}\sin\left(\min\left((k+1)\alpha+\varphi_0,\frac{\pi}{2}\right)\right)} \right.\nonumber \\
&&\qquad  \left. -\erf{\sqrt{2y}\sin\left(\min\left((k+1)\alpha-\varphi_0,\frac{\pi}{2}\right)\right)} \right]  \nonumber \\
&&  \kern-1em +\left(\frac{2}{\pi}\right)^{3/2} \sqrt{y} \; \frac{\mathrm{e}^{-y}}{2} \int_0^{+\infty} \mathrm{d}u \; \mathrm{e}^{-y\cosh u} \sinh\frac{u}{2} \times \nonumber \\
&&\kern-1em  \left[ \arctan\left( \frac{\sin\left(\frac{\pi}{\alpha}\left(\varphi_0+\frac{\pi}{2}\right)\right)}{\sinh\left(\frac{\pi u}{2\alpha}\right)}\right) +\arctan\left( \frac{\sin\left(\frac{\pi}{\alpha}\left(\varphi_0-\frac{\pi}{2}\right)\right)}{\sinh\left(\frac{\pi u}{2\alpha}\right)}\right)  \right] \nonumber \\
\end{eqnarray}
and in an obtuse wedge of top angle $\alpha \geqslant \pi$
\begin{eqnarray}\label{obtuse}
&&\kern-1emS(y,\varphi_0) = \mathrm{erf}\left(\sqrt{2y}\sin\left(\min\left(\varphi_0,\frac{\pi}{2}\right) \right)\right) \nonumber \\
&&\kern-1em+\left(\frac{2}{\pi}\right)^{3/2} \sqrt{y} \frac{\mathrm{e}^{-y}}{2} \int_0^{+\infty} \mathrm{d}u \; \mathrm{e}^{-y\cosh u} \sinh\frac{u}{2}  \times \nonumber \\
&& \kern-1em \left[ \arctan\kern-0.2em \left( \frac{\sin\left(\frac{\pi}{\alpha}\left(\varphi_0+\frac{\pi}{2}\right)\right)}{\sinh\left(\frac{\pi u}{2\alpha}\right)}\right) \kern-0.3em+\arctan\kern-0.2em\left( \frac{\sin\left(\frac{\pi}{\alpha}\left(\varphi_0-\frac{\pi}{2}\right)\right)}{\sinh\left(\frac{\pi u}{2\alpha}\right)}\right) \kern-0.2em \right]\nonumber \\
\end{eqnarray}
with $k=\lfloor\pi/(2\alpha) -1/2\rfloor$ and $y=r_0^2/(8Dt)$.
We condense Eqs.~\eqref{acute} and \eqref{obtuse} into the following expression
\begin{eqnarray}\label{condensed}
&&\kern-1em S(y,\varphi_0) = \erf{\sqrt{2y}\sin\left(\min\left(\varphi_0,\frac{\pi}{2}\right) \right)}\nonumber \\
&& +\sum\limits_{j=1}^{k+1} (-1)^j \left[\erf{\sqrt{2y}\sin\left(\min\left(j\alpha+\varphi_0,\frac{\pi}{2}\right)\right)} \right.\nonumber \\
&&\qquad  \left. -\erf{\sqrt{2y}\sin\left(\min\left(j\alpha-\varphi_0,\frac{\pi}{2}\right)\right)} \right]  \nonumber \\
&&  \kern-1em +\left(\frac{2}{\pi}\right)^{3/2} \sqrt{y} \; \frac{\mathrm{e}^{-y}}{2} \int_0^{+\infty} \mathrm{d}u \; \mathrm{e}^{-y\cosh u} \sinh\frac{u}{2} \times \nonumber \\
&&\kern-1em  \left[ \arctan\kern-0.2em \left( \frac{\sin\left(\frac{\pi}{\alpha}\left(\varphi_0+\frac{\pi}{2}\right)\right)}{\sinh\left(\frac{\pi u}{2\alpha}\right)}\right) \kern-0.3em+\arctan\kern-0.2em\left( \frac{\sin\left(\frac{\pi}{\alpha}\left(\varphi_0-\frac{\pi}{2}\right)\right)}{\sinh\left(\frac{\pi u}{2\alpha}\right)}\right) \kern-0.2em \right] \kern-0.2em.\nonumber \\
\end{eqnarray}
 This expression is valid for any arbitrary wedge angle and generalizes the expression of Dy and Esguerra that requires wedge angles of the form $\pi/n$ with $n$ an integer \cite{Dy:2008}, as we proceed to show in the next paragraph. Note that the case of obtuse wedges was not covered by their approach.

\subsubsection{Particular cases $\pi/n$}

We check that in the particular cases of wedge angles of the form $\pi/n$ with $n$ an integer, we recover the expressions of Dy and Esguerra \cite{Dy:2008,Dy:2013}.
 First, if \mbox{$n=2p+1$}, the integral part of the survival probability disappears, as easily seen in Eq.~\eqref{survieydev}. In this case, $k=p$, and
\begin{eqnarray}
&&\min\left((k+1)\alpha\pm\varphi_0,\frac{\pi}{2}\right)=\frac{\pi}{2} \nonumber
\end{eqnarray}
so the expression becomes
\begin{eqnarray}
&&S(y,\varphi_0) = \erf{\sqrt{2y}\sin\varphi_0} \nonumber \\
&& \quad +\sum\limits_{j=1}^{p} (-1)^j   \left[\erf{\sqrt{2y}\sin(j\alpha+\varphi_0)} \right. \nonumber \\
&& \qquad \qquad \qquad \left.-\erf{\sqrt{2y}\sin(j\alpha-\varphi_0)} \right] \label{nous} \\
&&  = \sum_{k=0}^{n-1} \frac{(-1)^k}{4} \left[ \erf{\sqrt{2y}\sin(k\alpha+\varphi_0)} \right. \nonumber \\
&& \qquad \qquad +\erf{\sqrt{2y}\sin(-k\alpha+\varphi_0)} \nonumber \\
&&\qquad \qquad +\erf{\sqrt{2y}\sin((k+1)\alpha-\varphi_0)} \nonumber \\
&& \qquad \qquad \left. +\erf{\sqrt{2y}\sin((1-k)\alpha-\varphi_0)} \right]. \label{Dy}
\end{eqnarray}
Equation \eqref{nous} matches the one of Dy and Esguerra \cite{Dy:2008} reminded in Eq.~\eqref{Dy}.

Then, if the wedge angle is now $\pi/n$ with $n=2p$, $k=p-1$, so
\begin{eqnarray}
&&\min\left((k+1)\alpha+\varphi_0,\frac{\pi}{2}\right)=\frac{\pi}{2} \nonumber \\
&&\min\left((k+1)\alpha-\varphi_0,\frac{\pi}{2}\right) =(k+1)\alpha-\varphi_0=\frac{\pi}{2}-\varphi_0 \nonumber
\end{eqnarray}
and the survival probability can be rewritten
\begin{eqnarray}
&&S(y,\varphi_0) = \mathrm{erf}\left(\sqrt{2y}\sin\varphi_0\right) \nonumber \\
&&  +\sum\limits_{j=1}^{p-1} (-1)^j   \left[\mathrm{erf}\left(\sqrt{2y}\sin(j\alpha+\varphi_0)\right)  \right. \nonumber \\
&& \quad \qquad \qquad \quad \left . - \, \mathrm{erf}\left(\sqrt{2y}\sin(j\alpha-\varphi_0)\right) \right]  \nonumber \\
&&+(-1)^{p}  \left[ \mathrm{erf}\left(\sqrt{2y}\right) -\, \mathrm{erf}\left(\sqrt{2y}\cos\varphi_0 \right)\right] \nonumber \\
&& +(-1)^p \left(\frac{2}{\pi}\right)^{3/2} \sqrt{y} \; \mathrm{e}^{-y}  \times \nonumber \\
&&  \int_0^{+\infty} \kern-0.7em \mathrm{d}u \; \mathrm{e}^{-y\cosh u} \sinh\frac{u}{2}  \arctan\left( \frac{\sin\left( 2p \, \varphi_0\right)}{\sinh\left(p u\right)}\right)\kern-0.3em. 
\end{eqnarray}
This expression can be numerically checked to match the known expression of the survival probability for $n=2$ \cite{Dy:2008}
\begin{equation}
S(y,\varphi_0)=\mathrm{erf}(\sqrt{2y}\sin\varphi_0) \, \mathrm{erf}(\sqrt{2y}\cos\varphi_0).
\end{equation}
The next values $n=4$ and $n=6$ have been checked on the first-passage time distribution \cite{Dy:2013} (see Eq.~\eqref{FPT}).

\subsubsection{Discussion}

\begin{figure}[h]
\centering
\includegraphics[width=230pt]{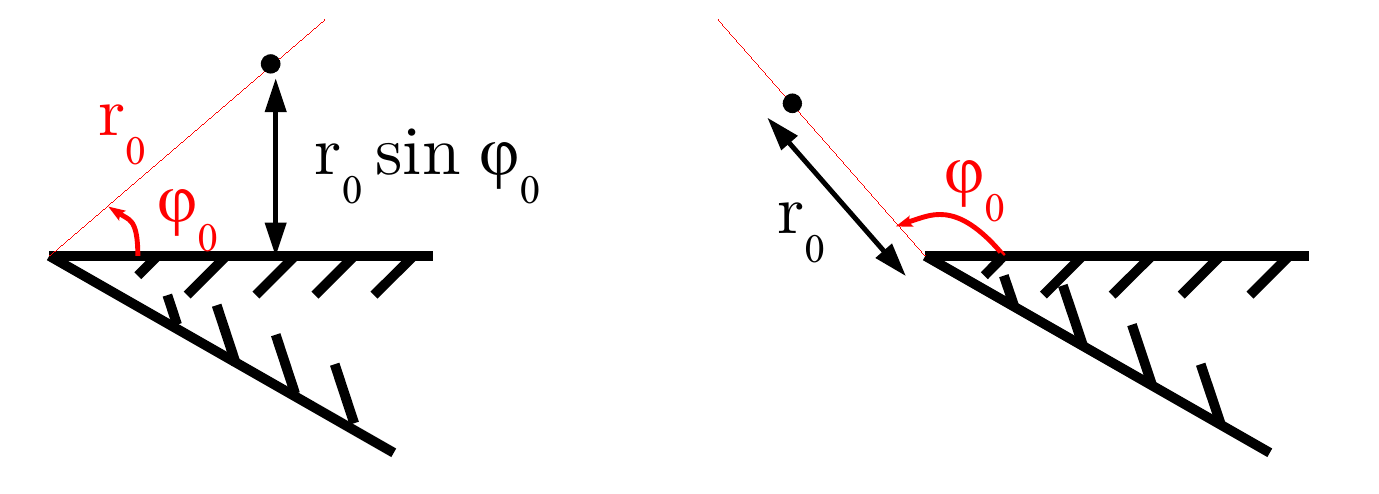}
\caption{Distance between the starting point and the wedge. If \mbox{$\varphi_0 \leqslant \pi/2$} (left), the projection of the starting point on the wedge is on the absorbing edge and the distance is \mbox{$r_0 \sin\varphi_0$}. If $\varphi_0 \leqslant \pi/2$, the projection of the starting point on the wedge is the apex and the distance is $r_0$.}
\label{distwedge}
\end{figure}

We make here several comments on the different terms involved in Eqs.~\eqref{acute} and \eqref{obtuse} (summarized in Eq.~\eqref{condensed}). Without loss of generality, we assume that the starting point is in the inferior part of the wedge (\mbox{$\varphi_0\leqslant \alpha/2$}). The first term of Eq.~\eqref{condensed} is the survival probability in an infinite half space delimited by an absorbing infinite plane \cite{Redner} for a walk whose starting distance from the plane is given by \mbox{$r_0 \sin (\mathrm{min} (\varphi_0,\pi/2))$}. This latter corresponds to the distance of the starting point of the walk to the closest absorbing boundary of the wedge (which is the one at $\varphi=0$ because of the choice \mbox{$\varphi_0\leqslant \alpha/2$}) in the original problem. If the wedge is acute, it is \mbox{$r_0 \sin\varphi_0$}. If the wedge is obtuse, this distance depends on whether the projection of the starting point on the axis $\varphi=0$ is on the absorbing edge or not (see Fig.~\ref{distwedge}). In the first case (corresponding to $\varphi_0 \leqslant \pi/2$), this distance is, like in the acute case, \mbox{$r_0 \sin\varphi_0$}. In the second case (for $\varphi_0 \geqslant \pi/2$), the distance is $r_0$, which is the distance from the starting point to the apex of the wedge.

Then, by comparison with the expressions of Dy and Esguerra \cite{Dy:2008,Dy:2013}, obtained in the particular case of $\pi/n$ wedge angles, the sum involved in Eq.~\eqref{acute} can be seen as a sum over generalized images (sinks and sources), that only exists for an acute wedge.

Finally, the integral term and the term $k+1$ of the sum of Eq.~\eqref{condensed} are the hallmark of a wedge angle different from \mbox{$\pi/(2p+1)$} with $p$ an integer.

\subsection{First-passage time distribution}

Similar expressions for the first-passage time distribution are easily obtained from Eq.~\eqref{condensed}. Knowing that
\begin{equation}
F(t)=-\deriv{S}{t},
\end{equation}
it is found that, for any planar wedge,
\begin{eqnarray}\label{FPT}
&&\kern-1em F(t)=\frac{r_0}{2\sqrt{\pi Dt^3}}  \Bigg\{ \sin\left(\min\left(\varphi_0,\frac{\pi}{2}\right)\right) \e{-\frac{r_0^2 \sin^2\left(\min\left(\varphi_0,\pi/2\right)\right) }{4Dt}}  \nonumber \\
&&+\sum\limits_{j=1}^{k+1} (-1)^j \times \nonumber \\
&&\kern-1em \left[ \sin\left(\min\left(j\alpha+\varphi_0,\frac{\pi}{2}\right)\right) \e{-\frac{r_0^2 \sin^2\left(\min\left(j\alpha+\varphi_0,\pi/2\right)\right)}{4Dt}}  \right. \nonumber \\
&&    \left.  -\sin\left(\min\left(j\alpha-\varphi_0,\frac{\pi}{2}\right)\right) \e{-\frac{r_0^2 \sin^2\left(\min\left(j\alpha-\varphi_0,\pi/2\right)\right)}{4Dt}}  \right] \nonumber \\
&& \kern-1em + \frac{1}{2\pi} \int_0^{+\infty} \diff{v} \left(1-\frac{r_0^2}{2Dt}\cosh^2 \frac{v}{2}\right) \sinh \frac{v}{2} \e{-\frac{r_0^2}{4Dt} \cosh^2\frac{v}{2}}  \times\nonumber \\
&& \kern-1em \left[ \arctan\kern-0.3em\left( \frac{\sin\left(\frac{\pi}{\alpha}\left(\varphi_0+\frac{\pi}{2}\right)\right)}{\sinh\left(\frac{\pi v}{2\alpha}\right)}\right) \kern-0.3em+\arctan\kern-0.2em\left( \frac{\sin\left(\frac{\pi}{\alpha}\left(\varphi_0-\frac{\pi}{2}\right)\right)}{\sinh\left(\frac{\pi v}{2\alpha}\right)}\right)  \kern-0.2em\right] \kern-0.5em\Bigg\}  \nonumber \\
\end{eqnarray}
with $k=\lfloor\pi/(2\alpha) -1/2\rfloor$ and $y=r_0^2/(8Dt)$.

\section{Asymptotic development of the survival probability at short times}

As an application of the previous results, we now show that the asymptotic development of the survival probability at short times ($y \gg 1$) can be conveniently extracted from Eqs.~\eqref{acute} and \eqref{obtuse}. The leading order is 1, as expected because a walker starting from the bulk cannot be on the absorbing boundaries at time $t=0$. We are interested in the corrections induced by the boundaries, and first address the case of an acute wedge.

Sine being a growing function in $\interval{0,\pi/2}$,
\begin{equation}\label{ineq}
\sin\varphi_0 \leqslant \sin(\alpha-\varphi_0) \leqslant \sin(\alpha+\varphi_0) \leqslant \sin(2\alpha-\varphi_0) \leqslant ... \leqslant 1.
\end{equation}
Moreover, the error function asymptotically grows like
\begin{equation}\label{asympt}
\erf{x} \equi{x \to \infty} 1-\e{-x^2} \mathrm{P}\left(\frac{1}{x}\right),
\end{equation}
with $\mathrm{P}$ a polynomial. The term $\erf{\sqrt{2y}\sin\varphi_0}$ of Eq.~\eqref{acute} thus contains the leading order of the survival probability and a first correction to this value, which turns out to be the main one. The successive terms of the sum involved in Eq.~\eqref{acute} can be shown to be smaller and smaller corrections by using the inequalities (\ref{ineq}) and the asymptotic expansion \eqref{asympt}.
Last, we evaluate the large $y$ asymptotics of the integral term
\begin{eqnarray}
&&  \kern-1em \mathcal{I} \equiv \left(\frac{2}{\pi}\right)^{3/2} \sqrt{y} \; \frac{\mathrm{e}^{-y}}{2} \int_0^{+\infty} \mathrm{d}u \; \mathrm{e}^{-y\cosh u} \sinh\frac{u}{2} \times \nonumber \\
&&\kern-1em  \left[ \arctan \kern-0.2em \left( \frac{\sin\left(\frac{\pi}{\alpha}\left(\varphi_0+\frac{\pi}{2}\right)\right)}{\sinh\left(\frac{\pi u}{2\alpha}\right)}\kern-0.2em\right) \kern-0.2em+\arctan\kern-0.2em\left( \frac{\sin\left(\frac{\pi}{\alpha}\left(\varphi_0-\frac{\pi}{2}\right)\right)}{\sinh\left(\frac{\pi u}{2\alpha}\right)}\kern-0.2em\right)  \kern-0.2em\right]. \nonumber \\
\end{eqnarray}
 Using Laplace's method, we have to distinguish the two following cases: (i) when $\sin\left(\pi\left(\varphi_0+\pi/2\right)/\alpha\right)$ and $\sin\left(\pi\left(\varphi_0-\pi/2\right)/\alpha\right)$ have the same sign, and (ii) when they have opposite signs.
In the first case, we approximate the integral by
\begin{equation}
\mathcal{I} \sim \pm \frac{\e{-2y}}{\sqrt{2\pi y}}.
\end{equation}
In the second case, 
\begin{equation}
\mathcal{I}  \sim \frac{\e{-2y}}{4\alpha y} \left( \frac{1}{\sin\left(\frac{\pi}{\alpha}\left(\varphi_0+\frac{\pi}{2}\right)\right)} + \frac{1}{\sin\left(\frac{\pi}{\alpha}\left(\varphi_0-\frac{\pi}{2}\right)\right)}  \right). 
\end{equation}

In both cases, the leading order of the integral term is dominated by all the other terms of the sum of error functions. Finally, the short time asymptotics of the survival probability can be defined using a scale of functions based on complementary error functions
\begin{eqnarray}\label{asymptotics}
&&S(t|y,\varphi_0) \equi{y \to \infty} 1+\psi_1(y,\varphi_0) \nonumber \\
&& \quad +\sum\limits_{j=1}^{k} [\psi_{2j}(y,\varphi_0) +\psi_{2j+1}(y,\varphi_0)] +R_{2k+2}(y,\varphi_0) \nonumber \\
\end{eqnarray}
with for $j\leqslant k$  
\begin{eqnarray}\label{erfscale}
&\psi_1(y,\varphi_0)=&-\erfc{\sqrt{2y}\sin\varphi_0}\nonumber \\
& \psi_{2j}(y,\varphi_0)=&(-1)^j   \erfc{\sqrt{2y}\sin(j\alpha-\varphi_0)} \nonumber \\
& \psi_{2j+1}(y,\varphi_0)=&(-1)^{j+1}   \erfc{\sqrt{2y}\sin(j\alpha+\varphi_0)} \nonumber \\
\end{eqnarray}
and the remainder
\begin{eqnarray}\label{remainder}
&&R_{2k+2}(y,\varphi_0)=\nonumber \\
&&(-1)^{k+1}  \left[ \mathrm{erfc}\left(\sqrt{2y}\sin\left(\min\left((k+1)\alpha-\varphi_0,\frac{\pi}{2}\right)\right)\right) \right.\nonumber \\
&& \qquad    \left. -\, \mathrm{erfc}\left(\sqrt{2y}\sin\left(\min\left((k+1)\alpha+\varphi_0,\frac{\pi}{2}\right)\right)\right)\right] \nonumber \\
&&  +\left(\frac{2}{\pi}\right)^{3/2} \sqrt{y} \; \frac{\mathrm{e}^{-y}}{2} \int_0^{+\infty} \mathrm{d}u \; \mathrm{e}^{-y\cosh u} \sinh\frac{u}{2}  \nonumber \\
&& \quad \times \left[ \arctan\left( \frac{\sin\left(\frac{\pi}{\alpha}\left(\varphi_0+\frac{\pi}{2}\right)\right)}{\sinh\left(\frac{\pi u}{2\alpha}\right)}\right) \right.\nonumber \\
&& \qquad \qquad \left. +\arctan\left( \frac{\sin\left(\frac{\pi}{\alpha}\left(\varphi_0-\frac{\pi}{2}\right)\right)}{\sinh\left(\frac{\pi u}{2\alpha}\right)}\right)  \right] 
\end{eqnarray}
such that  \mbox{$\psi_i(y,\varphi_0)=o\left(\psi_{i-1}(y,\varphi_0)\right)$}, and \mbox{$R_{2k+2}(y,\varphi_0)=o\left(\psi_{2k+1}(y,\varphi_0)\right)$}.

\begin{figure}[h]
\centering
\includegraphics[width=220pt]{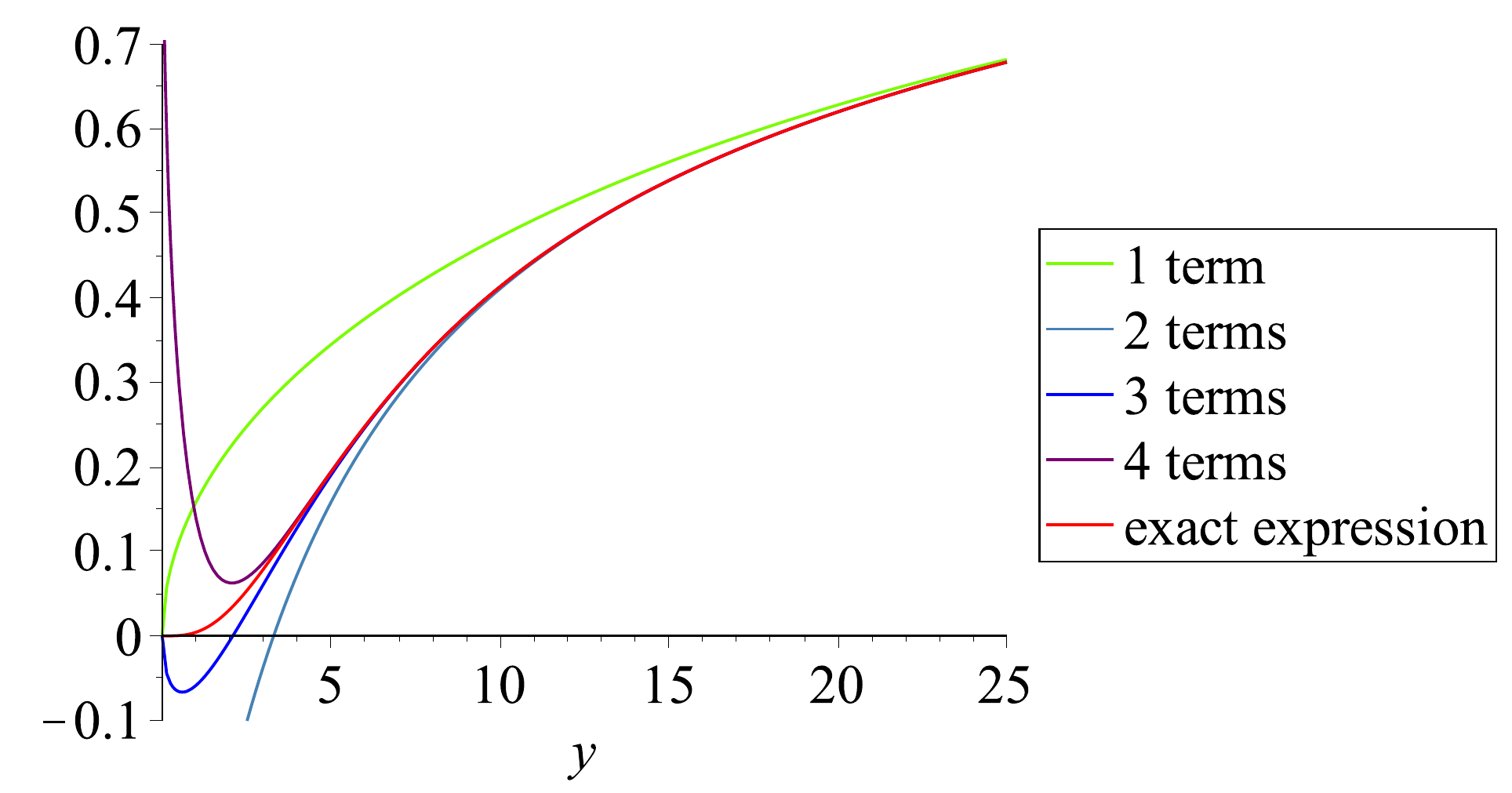}
\caption{Survival probability in a wedge of acute top angle $\alpha=0.4$ rad, with $\varphi_0=0.1$ rad (in red), and the short time development cut at different orders. The exact expression contains here 8 terms, including the remainder $R_8(y,\varphi_0)$. We can see that as soon as we keep 2 terms, the short time development has a very satisfying range of validity, that can be extended by keeping more terms in the development.}
\label{fig_acute}
\end{figure}

\begin{figure}[h]
\centering
\includegraphics[width=220pt]{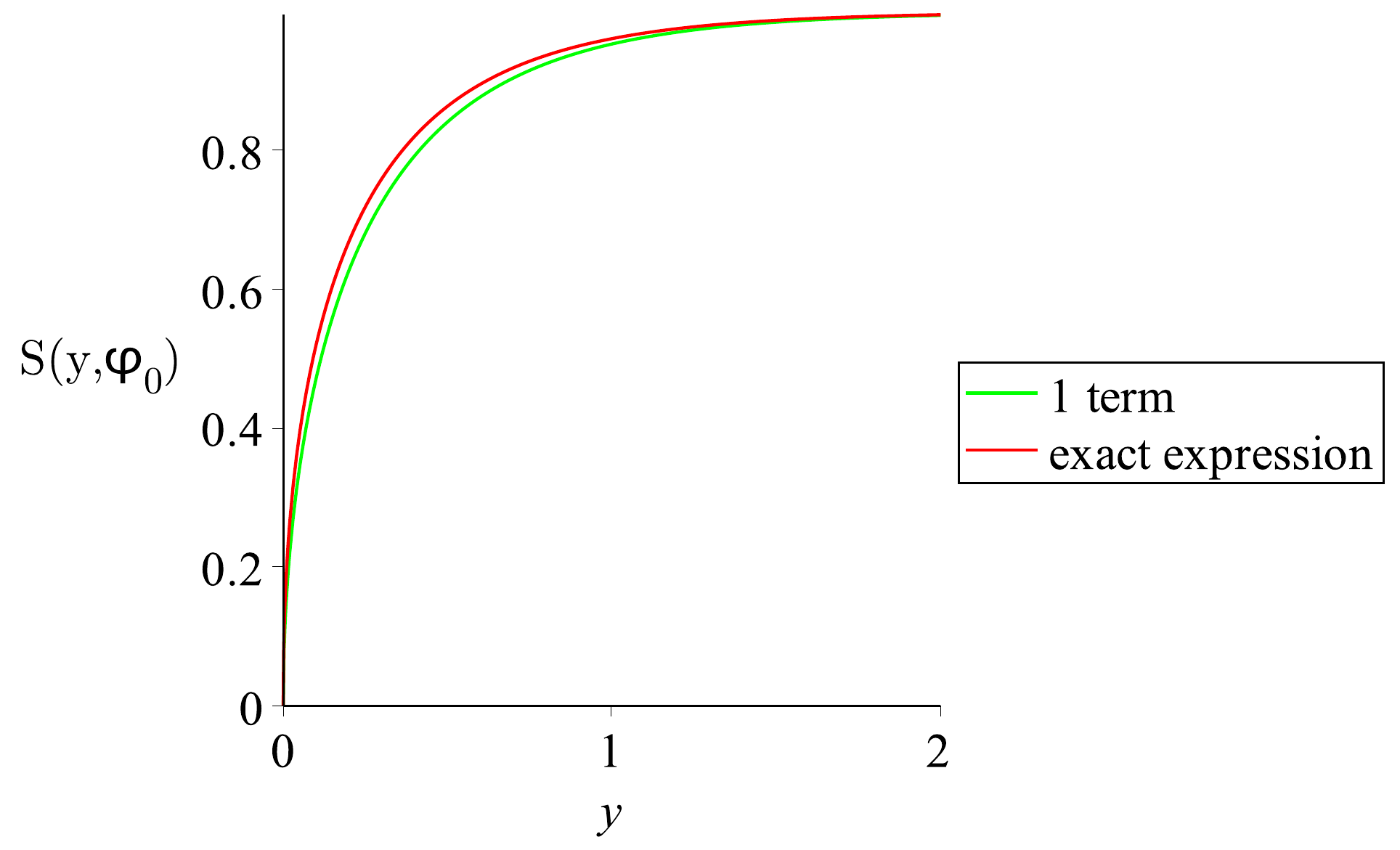}
\caption{Survival probability in an obtuse wedge in the case \mbox{$\varphi_0 \leqslant \pi/2$} (in red) and the term \mbox{$1-\erfc{\sqrt{2y}\sin \varphi_0}$} (in green). The top angle is $\alpha=3.5$ rad and $\varphi_0=1.5$ rad. The error function term provides an accurate approximation of the survival probability with a large range of validity.}
\label{fig_obtuse}
\end{figure}

The obtuse case is simpler. If \mbox{$\varphi_0 \leqslant \pi/2$}, the error function term gives the main correction and the integral term is subdominant, whereas if \mbox{$\varphi_0 \geqslant \pi/2$}, these two terms have the same exponential decay rate. In Fig.~\ref{fig_obtuse}, we only illustrate the first case.

The previous analysis shows that at short times (large $y$), for acute wedge angles and obtuse ones where \mbox{$\varphi_0 \leqslant \pi/2$}, the survival probability is mainly influenced by the edge closest to the starting point, producing the first correction to the limit value 1
\begin{equation}
S(y,\varphi_0) \equi{y\to\infty} 1- \erfc{\sqrt{2y} \sin \varphi_0}.
\end{equation}
Moreover, for acute wedges, our approach also gives a set of smaller and smaller corrections, the least correction being given by the remainder \mbox{$R_{2k+2}(y,\varphi_0)$}. We check on Figs.~\ref{fig_acute} and \ref{fig_obtuse} that the short time development is accurate and has a significant range of validity that increases, in the case of acute angles, with the number of correcting terms taken into account. In practice, it means that unless we need to describe the very long times (small $y$), the integral term, which is the most complicated to compute, can be dropped.

\section{Conclusion}

In this paper, we established \textit{simple} expressions of the survival probability and the first-passage time distribution in a planar wedge with infinite absorbing boundaries. The result holds for \textit{any} top angle of the wedge, and in particular covers the case of obtuse wedges. It thus generalizes the expressions obtained by Dy and Esguerra \cite{Dy:2008,Dy:2013}, which were limited to wedge angles of the form $\pi/n$ with $n$ a positive integer. The final expression only involves a finite sum of error functions, that can be seen as a sum over generalized images, and an integral of elementary functions. \\
The expression given here naturally displays a development of the survival probability at short times, whereas the standard form of the survival probability, that is written as an infinite sum of special functions, does not allow to get this expansion. Moreover, this short time development has a large range of validity.\\
 The case of biased diffusion in an arbitrary wedge (considered in \cite{Dy:2013} for specific angles of the form $\pi/n$) would be a natural extension of the formalism developed in this work.

OB acknowledges support from European Research Council starting Grant No. FPTOpt-277998. SNM acknowledges support by ANR grant
2011-BS04-013-01 WALKMAT.

\appendix
\section{Calculation of the integrals involved in the term $A_1$}
We give here details of the calculation of the term $A_1$, which is the sum of the four integrals $C_1$, \mbox{$C_2\equiv C_2^++C_2^-$}, $C_3^-$ and $C_3^+$. The first one is
\begin{eqnarray}
&&C_1=2 \int_0^{\alpha} \mathrm{d} \psi \; \mathrm{e}^{y\cos\psi} \cos\frac{\psi}{2}\int_0^{\frac{\pi \varphi_0}{\alpha}} \mathrm{d}x' \nonumber \\
&& \quad  \times \sum\limits_{m=0}^{+\infty} \cos((2m+1)x') \cos\left((2m+1)\frac{\pi\psi}{2\alpha}\right),
\end{eqnarray}
and using Eq.~\eqref{dirac} of the main text,
\begin{eqnarray}
&C_1=& 2 \int_0^{\alpha} \mathrm{d} \psi \; \mathrm{e}^{y\cos\psi} \cos\frac{\psi}{2} \int_0^{\frac{\pi \varphi_0}{\alpha}} \mathrm{d}x' \; \frac{\pi}{4} \delta\left(x'-\frac{\pi\psi}{2\alpha}\right). \nonumber \\
\end{eqnarray}
The integral over $x'$ is $\pi/4$ if \mbox{$\pi\psi/(2\alpha) \in [0,\pi\varphi_0/\alpha]$}, \textit{i.e.} if \mbox{$\psi \in [0,2\varphi_0]$}, and 0 otherwise. Thus, as we have by definition \mbox{$2\varphi_0 \leqslant \alpha$}, the part \mbox{$\psi \in \interval{2\varphi_0,\alpha}$} gives $0$ and
\begin{eqnarray}
&C_1 &=\frac{\pi}{2} \int_0^{2\varphi_0} \mathrm{d} \psi \; \mathrm{e}^{y\cos\psi} \cos\frac{\psi}{2}.
\end{eqnarray}
We change the variables $u=\sqrt{2 y} \sin(\psi/2)$ and get
\begin{eqnarray}\label{C1}
&C_1&=\left(\frac{\pi}{2}\right)^{3/2} \frac{\mathrm{e}^y}{\sqrt{y}} \; \mathrm{erf}(\sqrt{2y}\sin\varphi_0).
\end{eqnarray}

The second integral $C_2$ is the sum of $C_2^+$ and $C_2^-$, given by
\begin{eqnarray}
&C_2^{\pm}&=2 \sum\limits_{j=1}^{k} \int_0^{\alpha} \mathrm{d}\psi \; \mathrm{e}^{y\cos(2j\alpha\pm\psi)} \cos\left(j\alpha\pm\frac{\psi}{2}\right) \int_0^{\frac{\pi \varphi_0}{\alpha}} \mathrm{d}x' \nonumber \\
&& \kern-1em  \times \sum\limits_{m=0}^{+\infty} \cos((2m+1)x') \cos\left((2m+1)j \pi\pm(2m+1)\frac{\pi\psi}{2\alpha}\right) \nonumber \\
&&\kern-1em=2 \sum\limits_{j=1}^{k} (-1)^j \int_0^{\alpha} \mathrm{d}\psi \; \mathrm{e}^{y\cos(2j\alpha\pm\psi)} \cos\left(j\alpha\pm\frac{\psi}{2}\right) \nonumber \\
&&\kern-1em \quad \times \int_0^{\frac{\pi \varphi_0}{\alpha}} \mathrm{d}x' \sum\limits_{m=0}^{+\infty} \cos((2m+1)x') \cos\left((2m+1)\frac{\pi\psi}{2\alpha}\right) \nonumber \\
\end{eqnarray}
Following the same lines, we obtain
\begin{eqnarray}\label{C2}
&&C_2=C_2^-+C_2^+=\sum\limits_{j=1}^{k} (-1)^j \left(\frac{\pi}{2}\right)^{3/2} \frac{\mathrm{e}^y}{\sqrt{y}} \nonumber \\
&& \quad \times \left[\mathrm{erf}(\sqrt{2y}\sin(j\alpha+\varphi_0))-\mathrm{erf}(\sqrt{2y}\sin(j\alpha-\varphi_0))\right]. \nonumber \\
\end{eqnarray}

Finally, we compute carefully the last two integrals, because of their integration limits
\begin{eqnarray}
&&C_3^-=(-1)^{k+1} \frac{\pi}{2}\int_{\max(0,(2k+2)\alpha-\pi)}^{\alpha} \mathrm{d}\psi \; \mathrm{e}^{y\cos((2k+2)\alpha-\psi)} \nonumber \\
&& \quad \times \cos\left((k+1)\alpha-\frac{\psi}{2}\right)  \int_0^{\frac{\pi \varphi_0}{\alpha}} \mathrm{d}x \; \delta\left(x-\frac{\pi\psi}{2\alpha}\right)
\end{eqnarray}
As previously, the integral over $x$ equals to $1$ if \mbox{$\psi\in\interval{0,2\varphi_0}$}, and $0$ otherwise. If the inferior limit is larger than $2\varphi_0$, the integral is $0$. We can then rewrite it as
\begin{eqnarray}
&&C_3^-=(-1)^{k+1} \frac{\pi}{2}\int_{\max(0,(2k+2)\alpha-\pi)}^{\max(2\varphi_0,\max(0,(2k+2)\alpha-\pi))} \mathrm{d}\psi \nonumber \\
&& \qquad \times \; \mathrm{e}^{y\cos((2k+2)\alpha-\psi)} \cos\left((k+1)\alpha-\frac{\psi}{2}\right).
\end{eqnarray}
As $\varphi_0\geqslant 0$, we notice that
\begin{equation}
\max(2\varphi_0,\max(0,(2k+2)\alpha-\pi))=\max(2\varphi_0,(2k+2)\alpha-\pi).
\end{equation}
Moreover, as \mbox{$-\max(a,b)=\min(-a,-b)$}, we obtain
\begin{eqnarray}\label{C3-}
&&\kern-2em  C_3^-=(-1)^{k+1} \left(\frac{\pi}{2}\right)^{3/2} \frac{\mathrm{e}^y}{\sqrt{y}} \nonumber \\
&&\kern-1em \times \left[ \mathrm{erf}\left(\sqrt{2y}\sin\left(\min\left((k+1)\alpha,\frac{\pi}{2}\right)\right)\right) \right.\nonumber \\
&&\kern-1em \quad \left. - \, \mathrm{erf}\left(\sqrt{2y}\sin\left(\min\left((k+1)\alpha-\varphi_0,\frac{\pi}{2}\right)\right)\right)\right]. 
\end{eqnarray}

We proceed similarly for the last integral:
\begin{eqnarray}
&&C_3^+=(-1)^{k+1} \frac{\pi}{2}\int_0^{\max(0,\pi-(2k+2)\alpha)} \mathrm{d}\psi \; \mathrm{e}^{y\cos((2k+2)\alpha+\psi)} \nonumber \\
&& \quad \times \cos\left((k+1)\alpha+\frac{\psi}{2}\right)  \int_0^{\frac{\pi \varphi_0}{\alpha}} \mathrm{d}x' \; \delta\left(x'-\frac{\pi\psi}{2\alpha}\right) 
\end{eqnarray}
which is $0$ if \mbox{$\max(0,\pi-(2k+2)\alpha)=0$}. We can then rewrite
\begin{eqnarray}
&&C_3^+=(-1)^{k+1} \frac{\pi}{2}\int_{\min(0,\pi-(2k+2)\alpha)}^{\pi-(2k+2)\alpha} \mathrm{d}\psi \; \mathrm{e}^{y\cos((2k+2)\alpha+\psi)} \nonumber \\
&& \quad \times \cos\left((k+1)\alpha+\frac{\psi}{2}\right)  \int_0^{\frac{\pi \varphi_0}{\alpha}} \mathrm{d}x' \; \delta\left(x'-\frac{\pi\psi}{2\alpha}\right) 
\end{eqnarray}
and end the calculation as before to get
\begin{eqnarray}\label{C3+}
&&\kern-2em  C_3^-=(-1)^{k+1} \left(\frac{\pi}{2}\right)^{3/2} \frac{\mathrm{e}^y}{\sqrt{y}} \nonumber \\
&&\kern-1em \times \left[  \mathrm{erf}\left(\sqrt{2y}\sin\left(\min\left((k+1)\alpha+\varphi_0,\frac{\pi}{2}\right)\right)\right)\right.\nonumber \\
&&\kern-1em \quad \left. - \, \mathrm{erf}\left(\sqrt{2y}\sin\left(\min\left((k+1)\alpha,\frac{\pi}{2}\right)\right)\right)\right]. 
\end{eqnarray}
 Summing Eqs.~\eqref{C1}, \eqref{C2}, \eqref{C3-} and \eqref{C3+} leads to Eq.~\eqref{A1} of the main text.


\begin{thebibliography}{10}

\bibitem{Redner}
S.~Redner, {\em A guide to first-passage processes}.
\newblock Cambridge University Press, 2001.

\bibitem{Majumdar:1999a}
S.~N. Majumdar, ``Persistence in nonequilibrium systems,'' {\em Curr. Sci.},
  vol.~77, p.~370, 1999.

\bibitem{Bray:2013}
A.~J. Bray, S.~N. Majumdar, and G.~Schehr, ``Persistence and first-passage
  properties in nonequilibrium systems,'' {\em Advances in Physics}, vol.~62,
  pp.~225--361, 2013/07/04 2013.

\bibitem{BenichouNatChem10}
O.~B{\'e}nichou, C.~Chevalier, J.~Klafter, B.~Meyer, and R.~Voituriez,
  ``{Geometry-controlled kinetics.},'' {\em {Nat Chem}}, vol.~2, pp.~472--7,
  June 2010.

\bibitem{Meyer:2011}
B.~Meyer, C.~Chevalier, R.~Voituriez, and O.~B{\'e}nichou, ``Universality
  classes of first-passage-time distribution in confined media,'' {\em Physical
  Review E}, vol.~83, no.~5, p.~051116, 2011.

\bibitem{CondaminNature07}
S.~Condamin, O.~B{\'e}nichou, V.~Tejedor, R.~Voituriez, and J.~Klafter,
  ``{First-passage times in complex scale-invariant media.},'' {\em {Nature}},
  vol.~450, pp.~77--80, Nov. 2007.

\bibitem{Hughes}
B.~D. Hughes, {\em Random walks and random environments}.
\newblock Clarendon Press Oxford, 1996.

\bibitem{Hilhorst91}
M.~J. A.~M. Brummelhuis and H.~J. Hilhorst, ``{Covering of a finite lattice by
  a random walk},'' {\em Physica A: Statistical and Theoretical Physics},
  vol.~176, pp.~387--408, Sept. 1991.

\bibitem{Majumdar:2010}
S.~N. Majumdar, A.~Comtet, and J.~Randon-Furling, ``Random convex hulls and
  extreme value statistics,'' {\em Journal of Statistical Physics}, vol.~138,
  no.~6, pp.~955--1009, 2010.

\bibitem{Desbois:2003}
A.~Comtet and J.~Desbois, ``Brownian motion in wedges, last passage time and
  the second arc-sine law,'' {\em Journal of Physics A: Mathematical and
  General}, vol.~36, no.~17, p.~L255, 2003.

\bibitem{Lenzi:2009}
E.~Lenzi, ``Fokker-planck equation in a wedge domain: Anomalous diffusion and
  survival probability,'' {\em Physical Review E}, vol.~80, no.~2, 2009.

\bibitem{Metzler:2011}
J.-H. Jeon, A.~Chechkin, and R.~Metzler, ``First passage behaviour of
  fractional brownian motion in two-dimensional wedge domains,'' {\em EPL
  (Europhysics Letters)}, vol.~94, no.~2, p.~20008, 2011.

\bibitem{Lagache:2008}
T.~Lagache and D.~Holcman, ``Effective motion of a virus trafficking inside a
  biological cell,'' {\em SIAM Journal on Applied Mathematics}, vol.~68, no.~4,
  pp.~1146--1167, 2008.

\bibitem{Redner:1999}
S.~Redner and P.L.~Krapivsky, ``Capture of the lamb: Diffusing predators seeking
  a diffusing prey,'' {\em American Journal of Physics}, vol.~67, no.~12,
  pp.~1277--1283, 1999.

\bibitem{Fisher:1988}
M.~E. Fisher and M.~P. Gelfand, ``The reunions of three dissimilar vicious
  walkers,'' {\em Journal of statistical physics}, vol.~53, no.~1-2,
  pp.~175--189, 1988.

\bibitem{Dy:2008}
D.~Dy and J.~Esguerra, ``First-passage-time distribution for diffusion through
  a planar wedge,'' {\em Physical Review E}, vol.~78, no.~6, 2008.

\bibitem{Dy:2013}
D.~Dy and J.~Esguerra, ``First-passage characteristics of biased diffusion in a
  planar wedge,'' {\em Physical Review E}, vol.~88, no.~1, 2013.

\bibitem{Turkevich:1985}
L.~Turkevich, ``Occupancy-probability scaling in diffusion-limited
  aggregation,'' {\em Physical Review Letters}, vol.~55, no.~9, pp.~1026--1029,
  1985.

\end{thebibliography}
\end{document}